\documentstyle[12pt,epsfig,amssymb]{article}

\begin{document}

\begin{titlepage}

\begin{center}

{\Large \bf Critical values for a non-attractive lattice gas model}

\vspace{20mm}

{\large  J. Ricardo G. de Mendon\c{c}a}

\vspace{5mm}

\centerline{\parbox{170mm}{Departamento de F\'{\i}sica, Universidade
Federal de S\~{a}o Carlos, 13565-905 S\~{a}o Carlos, SP, Brazil}}

\vspace{20mm}

{\large Abstract}

\vspace{5mm}

\parbox{120mm}
{We investigate numerically the critical behaviour of a one-dimensional
non-attractive lattice gas model that is the continuous-time version of
the Domany-Kinzel cellular automaton in one of its parameter subspaces.
The model shows a phase transition in the directed percolation universality
class of critical behaviour, as expected.

\vspace{5mm}

{\noindent}Keywords: critical values, lattice gas, finite-size scaling,
                     Domany-Kinzel cellular automaton, directed percolation

\vspace{5mm}

{\noindent}PACS numbers: 02.50.Ga, 05.40.+j, 64.60.Ht}

\end{center}

\end{titlepage}

%% %% %% %% %% %% %% %% %% %% %% %% %% %% %% %% %% %% %% %% %% %% %% %% %% %% %%

The Domany-Kinzel (DK) probabilistic cellular automaton (PCA) \cite{dk}
is one of the most studied PCA in the physics literature, because it is
the most general left-right symmetric one-dimensional PCA, and has the
interesting property of having the mixed site-bond directed percolation
(DP) process on the square lattice as one of its instances. Its defining
rules are given in table \ref{TAB-I}. The mixed site-bond DP process is
given by assigning a probability $s \in [\,0,1]$ for a site to be present
in the lattice, and a probability $b \in [\,0,1]$ for a bond to exist between
any two sites on the lattice. The mixed DP problem consists in finding the
values of $s$ and $b$ for which an infinite cluster of sites connected by
the bonds occurs such that one can walk unidirectionally in it, say to the
south and to the east indefinitely. In the DK PCA this problem is obtained
by choosing the transition probabilities $x=0$, $y=sb$, and $z=sb(2-b)$.
For $s=1$ one obtains the pure bond DP problem, whilst for $b=1$ one obtains
the pure site DP problem.

\begin{table}[b]
\centering
\caption{Rule table for the DK PCA. The first line gives the initial
neighbourhood, the other two lines give the probability at which the
state listed at left is reached by the central bit.}
\label{TAB-I}
\medskip
\begin{tabular}{c@{ }cccccccc}
\hline \hline
   &  000  &  001  &  010  &  011  &  100  &  101  &  110  &  111  \\
\hline
0: & $1-x$ & $1-y$ & $1-x$ & $1-y$ & $1-y$ & $1-z$ & $1-y$ & $1-z$ \\
1: &  $x$  &  $y$  &  $x$  &  $y$  &  $y$  &  $z$  &  $y$  &  $z$  \\
\hline \hline
\end{tabular}
\end{table}

In this paper we investigate numerically the critical behaviour of a
continuous-time one-dimensional non-attractive lattice gas for which some
lower bounds on the critical points of the PCA version was given recently
\cite{konno}. The model is related with the DK PCA in one of its parameter
subspaces, and although the model is not on the mixed site-bond DP parameter
subspace of the DK PCA, it presents DP exponents, as expected on the basis
of the DP conjecture \cite{conj}.

Let $n_{\ell}(t) \in \{0,1\}$ denote the occupation number of the site
$\ell \in \Lambda$ at the integer instant $t$, with $\Lambda \subset {\Bbb Z}$
a finite lattice with $|\Lambda|=L$ sites and periodic boundary conditions
$\ell+L \equiv \ell$. The model we are interested in is the continuous-time
version of the PCA defined by the rules
\begin{equation}
\label{PCA}
n_{\ell}(t+1)=
\left\{\begin{array}{c@{ \quad}l}
(n_{\ell-1}(t)+n_{\ell+1}(t)) \bmod 2 & \mbox{with probability  $p$}, \\
                   0                  & \mbox{with probability $1-p$}.
       \end{array}\right.
\end{equation}
The rule table for this PCA is given in table \ref{TAB-II}. From tables
\ref{TAB-I} and \ref{TAB-II} we see that our PCA is equivalent to the DK
PCA with rates $x=0$, $y=p$, and $z=0$. We thus see that unless we take
the unphysical value $b=2$ in the site-bond DP subspace of the DK PCA, this
model does not belong to that subspace.

\begin{table}[t]
\centering
\caption{Same as table \ref{TAB-I} for the PCA defined by Eq.~(\ref{PCA}).}
\label{TAB-II}
\medskip
\begin{tabular}{c@{ }cccccccc}
\hline \hline
   & 000 &  001  & 010 &  011  &  100  & 101 &  110  & 111 \\
\hline
0: &  1  & $1-p$ &  1  & $1-p$ & $1-p$ &  1  & $1-p$ &  1  \\
1: &  0  &  $p$  &  0  &  $p$  &  $p$  &  0  &  $p$  &  0  \\
\hline \hline
\end{tabular}
\end{table}

Our approach in constructing the continuous-time version for the above PCA
is to take its non-diagonal transitions, i.e., those transitions for which
the final state differs from the initial state, and associate with them a
stochastic lattice gas with transition rates given by the original PCA rules.
This approach has been used before in the PCA literature \cite{rujan}, and
is equivalent to the so-called `Hamiltonian' or `strong anisotropic' limit
for the transfer matrixes of equilibrium lattice models \cite{kogut}.

As is well known \cite{master}, we may write the master
equation for interacting lattice gases as a Schr\"{o}dinger-like equation
in Euclidean time,
\begin{equation}
\frac{{\rm d}}{{\rm d}t}|P(t)\rangle = -H|P(t)\rangle,
\end{equation}
with $|P(t)\rangle$ the generating vector of the probabilities $P({\bf n},t)=
\langle{\bf n}|P(t)\rangle$ of observing the configuration ${\bf n}=(n_{1},
n_{2},\ldots,n_{L}) \in \{0,1\}^{\Lambda}$ at instant $t$, and with the
infinitesimal generator $H$ of the Markov semigroup playing the role of
the Hamiltonian. For the non-diagonal transitions of the PCA defined by
Eq.~(\ref{PCA}), the operator $H$ can be written as
\begin{equation}
H = -\sum_{\ell=1}^{L}H_{\ell-1,\ell,\ell+1},
\end{equation}
with the three-body stochastic transition matrix $H_{\ell-1,\ell,\ell+1}$
given by
\begin{equation}
\label{THREE}
H_{\ell-1,\ell,\ell+1}= \left(\begin{array}{cccccccc}
\cdot & \cdot & \ 1   & \cdot & \cdot & \cdot & \cdot & \cdot \\
\cdot &  -p   & \cdot & \ 1-p & \cdot & \cdot & \cdot & \cdot \\
\cdot & \cdot &  -1   & \cdot & \cdot & \cdot & \cdot & \cdot \\
\cdot & \ p   & \cdot &  -1+p & \cdot & \cdot & \cdot & \cdot \\
\cdot & \cdot & \cdot & \cdot &  -p   & \cdot & \ 1-p & \cdot \\
\cdot & \cdot & \cdot & \cdot & \cdot & \cdot & \cdot & \ 1   \\
\cdot & \cdot & \cdot & \cdot & \ p   & \cdot &  -1+p & \cdot \\
\cdot & \cdot & \cdot & \cdot & \cdot & \cdot & \cdot &  -1   
                        \end{array}\right),
\end{equation}
where the three-site basis vectors are ordered as usual, $(0,0,0) \prec
(0,0,1) \prec \cdots \prec (1,1,0) \prec (1,1,1)$, and the dots indicate
null entries. Proper tensorization of the above three-body matrix with unit
matrixes in order to obtain the full matrix $H$ is understood. Notice that
the elements in the columns of $H_{\ell-1,\ell,\ell+1}$ (and consequently
of $H$) add to zero due to the conservation of probabilities, and that its
non-diagonal elements are positive, since $0 \leq p \leq 1$. Identifying a
particle with the up spin state and a hole with the down spin state in the
$\sigma^{z}$ basis, the transition matrix $H_{\ell-1,\ell,\ell+1}$ above
is seen to be equivalent to the non-Hermitian quantum spin operator
\begin{equation}
H_{\ell-1,\ell,\ell+1}=
\frac{1}{2}(\sigma^{x}_{\ell}-1)\left[1+(1-p)\sigma^{z}_{\ell}+
p\sigma^{z}_{\ell-1}\sigma^{z}_{\ell}\sigma^{z}_{\ell+1}\right],
\end{equation}
where $\sigma^{x}$ and $\sigma^{z}$ are the usual Pauli spin-$\frac{1}{2}$
matrices. The transition matrix (\ref{THREE}) resembles the analogous matrix
for the basic contact process, but with non-standard rates and with the
elementary process $101 \to 111$ lacking. This lack is the root of the
non-attractiveness of the process. (Loosely speaking, attractive interacting
particle systems present a tendency for clustering, as it occurs in
ferromagnetic models or in the basic contact process. The precise mathematical
statement of attractiveness can be found in \cite{liggett}.)

The lowest gap in the spectrum of $H$ may be used to perform a finite-size
scaling analysis in the same way as one does in equilibrium problems
\cite{julia}. Around the critical point $p \gtrsim p^{*}$, the correlation
lengths of the infinite system behave like
\begin{equation}
\label{CORR}
\xi_{\|} \propto \xi_{\perp}^{z} \propto (p-p^{*})^{-\nu_{\|}}
                                 \propto (p-p^{*})^{-\nu_{\perp}z},
\end{equation}
where $\xi_{\|}$ and $\xi_{\perp}$ are the correlation lengths respectively
in the time and space directions, $\nu_{\|}$ and $\nu_{\perp}$ are the 
corresponding critical exponents, and  $z=\nu_{\|}/\nu_{\perp}$ is the
dynamical critical exponent. For finite systems of size $L$ we expect that
\begin{equation}
\label{CSI}
\xi_{\|,L}^{-1} = 
L^{-z_{L}}\Phi\left(|p-p^{*}_{L}|L^{1/\nu_{\perp,L}}\right),
\end{equation}
where $z_{L}$ and $\nu_{\perp,L}$ are the finite versions of $z$ and 
$\nu_{\perp}$, and $\Phi(u)$ is a scaling function with $\Phi(u \gg 1)
\sim u^{\nu_{\|}}$. On general grounds one expects $\lim_{L \to \infty}
p^{*}_{L},z_{L},\nu_{\perp,L}=p^{*},z,\nu_{\perp}$. From Eqs.~(\ref{CORR})
and (\ref{CSI}) we obtain
\begin{equation}
\label{PCTHETA}
      \frac{\ln \left[ \xi_{\|,L}(p^{*}_{L})/\xi_{\|,L'}(p^{*}_{L})
               \right]}{\ln (L/L')}  =
      \frac{\ln \left[ \xi_{\|,L''}(p^{*}_{L})/\xi_{\|,L}(p^{*}_{L})
                \right]}{\ln(L''/L)} =
      z_{L},
\end{equation}
which through the comparison of three different system sizes $L'< L < L''$ 
furnishes simultaneously $p^{*}_{L}$ and $z_{L}$. Of course, $\xi_{\|,L}$
and the gap $E_{L}^{(1)}-E_{L}^{(0)}=E_{L}^{(1)}$ of $H$ are related by
$\xi_{\|,L}^{-1}={\rm Re}\{E_{L}^{(1)}\}$.

We calculated the gaps of $H$ with the power method, which requires only
matrix-by-vector multiplications that can be carried out efficiently and
does not require a diagonalization in the usual, `QR' sense, a step that
may lessen the quality of the data. The version of the power method we use
takes advantage of the presence of absorbing states, and is also suitable for
the investigation of time dependent properties of Markov chains~\cite{power}.

\begin{figure}[t]
\centerline{\epsfig{figure=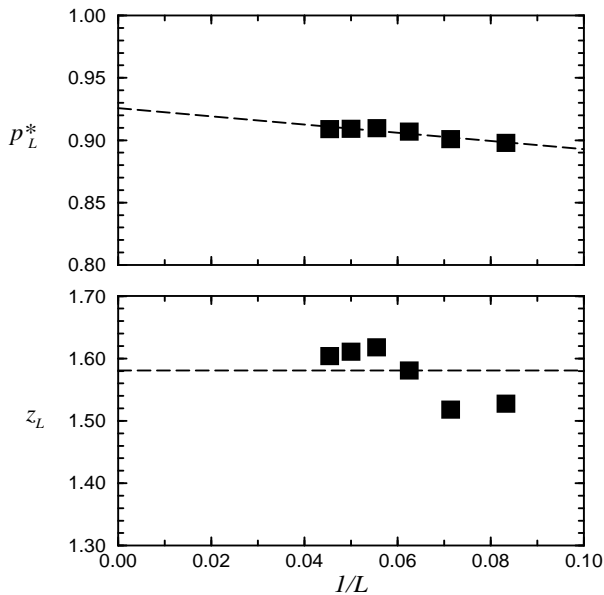,height=80mm,width=80mm}}
\caption{Finite-size data for the critical point $p^{*}$ and the dynamical
critical exponent $z$ of the model defined in Eq.~(\ref{THREE}). The dashed
line in the graph for $p^{*}_{L}$ is the least-squares linear fit to the data,
whilst the dashed line in the graph for $z_{L}$ represent the best known value
of $z_{\rm DP}$.}
\label{FIG}
\end{figure}

Our results for $p^{*}$ and $z$ are summarized in figure \ref{FIG}. Curiously
enough, despite the translational invariance of the lattice gas rules the
finite-size estimates behaved better for triplets of lengths of the form
$L',L,L''=2l-1,2l,2l+1$, $l \in {\Bbb N}$. Both sets of data behaved
irregularly with the system sizes, preventing us from applying the usual
extrapolation algorithms \cite{vbs,bst} to them. We are presumably in the
presence of strong finite-size effects and corrections to scaling. Notice
that the last three points of the data seem to be converging monotonically,
but unfortunately we were not able to go beyond $L=22$ in our diagonalizations.
The $L=\infty$ values for $p^{*}_{L}$ were obtained through a least-squares
fit to the curve $x_{L}=x_{\infty}+aL^{-1}$, since the data scale well with
$L^{-1}$ (although it does so with $L^{-2}$ also, but with a smaller
correlation coefficient), whilst for $z_{L}$ we only estimated the mean value
of our data. As expected on the basis of the DP conjecture, namely, that the
phase transition about a single absorbing state in single-component systems
with a scalar order parameter and in the absence of internal symmetries should
be in the DP universality class of critical behaviour \cite{conj}, our lattice
gas shows a DP-compatible exponent $z=1.58 \pm 0.04$. The most precise value
of $z$ for the DP universality class to date is given by $z=1.580\,745 \pm
0.000\,010$ \cite{jensen}. The critical point of the model is estimated as
$p^{*}=0.926 \pm 0.004$ (LS correlation coefficient $\gamma = -0.944$). This
value is slightly higher than the critical value $p_{\rm DK}^{*}=0.82 \pm
0.01$ of the corresponding point ($x=0$, $y=p$, $z=0$) in the DK PCA
\cite{dk}. This shift in the critical point for the lattice gas version of
the PCA was observed before in a study similar to the present one, where
the properties of the lattice gas on the line $x=0$, $y=z$ (corresponding
to the pure site DP problem) was investigated \cite{rujan}, and is probably
a general feature, since asynchronous dynamics tend to be more noisy than
synchronous dynamics.

In summary, we conducted numerical diagonalizations of the infinitesimal
generator of the continuous-time version of a non-attractive probabilistic
cellular automaton (PCA) that is an instance of the Domany-Kinzel (DK) PCA.
Although our PCA is not in the site-bond directed percolation (DP) subspace
of the DK PCA, its continuous-time version shows DP critical behaviour. Our
finite-size data showed an irregular approach to the infinite system limit,
and this prevented us from obtaining good estimates of the critical values.
It would be interesting to study this lattice gas by time-dependent Monte
Carlo methods in order to obtain more accurate critical values.

\medskip

This work was supported by the Funda\c{c}\~{a}o de Amparo \`{a} Pesquisa
do Estado de S\~{a}o Paulo (FAPESP), Brazil.

\end{document}